\documentclass[pdflatex,sn-mathphys-num]{sn-jnl}% Math and Physical Sciences Numbered Reference Style 
\usepackage{graphicx}%
\usepackage{multirow}%
\usepackage{amsmath,amssymb,amsfonts}%
\usepackage{amsthm}%
\usepackage{mathrsfs}%
\usepackage[title]{appendix}%
\usepackage{xcolor}%
\usepackage{textcomp}%
\usepackage{manyfoot}%
\usepackage{booktabs}%
\usepackage{algorithm}%
\usepackage{algorithmicx}%
\usepackage{algpseudocode}%
\usepackage{listings}%
\usepackage{amsmath}
\usepackage{mathtools}
\usepackage{bbm}
\usepackage{adjustbox}  
\usepackage{amssymb}
\usepackage{array}
\usepackage{multirow}
\usepackage{colortbl}  % 표 내 색상 적용 가능

\theoremstyle{thmstyleone}%
\theoremstyle{thmstyletwo}%

\theoremstyle{thmstylethree}%

\begin{document}

\title[Article Title]{Assessing the risk of recurrence in early-stage breast cancer through H\&E stained whole slide images }


\author*[1]{\fnm{Geongyu} \sur{Lee}}\email{gglee@deepbio.co.kr}

\author[1]{\fnm{Joonho} \sur{Lee}}\email{joonho.lee@deepbio.co.kr}

\author[1]{\fnm{Tae-Yeong} \sur{Kwak}}\email{tykwak@deepbio.co.kr}

\author[1]{\fnm{Sun Woo} \sur{Kim}}\email{swkim@deepbio.co.kr}

\author[2]{\fnm{Youngmee} \sur{Kwon}}\email{ymk@ncc.re.kr}

\author*[3]{\fnm{Chungyeul} \sur{Kim}}\email{idea1@korea.ac.kr}

\author*[4]{\fnm{Hyeyoon} \sur{Chang}}\email{hychang@deepbio.co.kr}

\affil[1]{\orgdiv{Research Team}, \orgname{Deep Bio Inc.}, \orgaddress{\street{Seoul}, \city{Seoul}, \postcode{08380}, \country{Korea Republic of}}}

\affil[2]{\orgdiv{Department of Pathology}, \orgname{National Cancer Center}, \orgaddress{\street{Goyang}, \city{Goyang}, \postcode{10408}, \country{Korea Republic of}}}

\affil[3]{\orgdiv{Department of Pathology}, \orgname{Korea University Guro Hospital}, \orgaddress{\street{Seoul}, \city{Seoul}, \postcode{08380}, \country{Korea Republic of}}}

\affil[4]{\orgdiv{Department of Pathology}, \orgname{Deep Bio Inc.}, \orgaddress{\street{Seoul}, \city{Seoul}, \postcode{08380}, \country{Korea Republic of}}}


\abstract{Accurate prediction of the likelihood of recurrence is important in the selection of postoperative treatment for patients with early-stage breast cancer. In this study, we investigated whether deep learning algorithms can predict patients' risk of recurrence by analyzing the pathology images of their cancer histology.We analyzed 125 hematoxylin and eosin-stained whole slide images (WSIs) from 125 patients across two institutions (National Cancer Center and Korea University Medical Center Guro Hospital) to predict breast cancer recurrence risk using deep learning. Sensitivity reached 0.857, 0.746, and 0.529 for low, intermediate, and high-risk categories, respectively, with specificity of 0.816, 0.803, and 0.972, and a Pearson correlation of 0.61 with histological grade. Class activation maps highlighted features like tubule formation and mitotic rate, suggesting a cost-effective approach to risk stratification, pending broader validation. These findings suggest that deep learning models trained exclusively on hematoxylin and eosin stained whole slide images can approximate genomic assay results, offering a cost-effective and scalable tool for breast cancer recurrence risk assessment. However, further validation using larger and more balanced datasets is needed to confirm the clinical applicability of our approach.}




\maketitle

\section{Introduction}\label{sec1}

Breast cancer is the most prevalent cancer worldwide, making early detection and rapid selection of treatment methods crucial for patients. \cite{giaquinto2022breast, siegel2023cancer} A significant number of early-stage breast cancers, particularly those that are estrogen receptor (ER)-positive and human epidermal growth factor receptor 2 (HER2)-negative, are classified as clinically low to intermediate risk based on traditional clinicopathological factors. This classification provides limited guidance for adjuvant chemotherapy decisions. Consequently, prognostic risk profiling has become a fundamental component of contemporary breast cancer diagnostics, offering additional risk information to identify patients who may not require adjuvant chemotherapy. \cite{giaquinto2022breast}
Among the various prognostic testing methods, genomic assays that assess molecular expression levels are noted for their relatively high predictive accuracy. \cite{howard2023integration, liu2022deep, rakha2010breast}  The Oncotype DX assay, in particular, evaluates the likelihood of recurrence in early breast cancer patients and the potential benefit from chemotherapy by analyzing the expression of 21 genes, including ER, PR, HER2 and Ki-67. This assay generates a Recurrence Score (RS) ranging from 0 to 100, categorizing risk into low, intermediate, and high. It is endorsed by major oncology guidelines and demonstrates high predictive reliability for breast cancer recurrence. \cite{paik2004multigene, sparano2018adjuvant} 
The role of traditional diagnostic parameters, such as tumor size, stage, and grade, remains vital in clinical decision-making. Histologic grade is an especially crucial prognostic factor in breast cancer. \cite{yin2020triple, darlix2019impact, zhao2020molecular} However, the classification of approximately 50\% of all breast cancers and around 60\% of ER-positive/HER2-negative tumors as histologic grade 2 introduces variability in aggressiveness and prognosis, limiting their utility in therapeutic decision-making.
The integration of digital pathology and the advancement of artificial intelligence, particularly deep learning, has enhanced the capabilities of computational pathology. \cite{mun2021yet, ryu2019automated, jung2022artificial, satturwar2024artificial} These technologies enable the prediction of patient outcomes, response to therapy, and molecular characteristics through the analysis of histopathological images, thus playing a pivotal role in the advancement of precision medicine. \cite{bentaieb2019deep, schneider2022integration, van2021deep, tolkach2020high}
This study introduced a novel approach to enhance prognosis prediction by employing deep learning to approximate traditional gene-based methods. By utilizing a self-supervised learning approach, specifically the contrastive loss function, the model was trained to identify morphological features common to each risk category, effectively removing uninformative data patches based on uncertainty. The activation maps generated confirmed that the models could replicate pathologists' assessment of morphological characteristics, such as mitotic figures. 
The principal finding of this research is noteworthy. The study verified that the developed model accurately predicts morphological features, including mitotic figures. A comparison of the model's predictions with the histological grading by pathologists showed a substantial agreement, proving the model's ability to replicate the assessments of expert pathologists effectively.
\section{Materials and Methods}

\subsection{Materials}
\begin{figure}[hp]
    \centering
    \includegraphics[width=260pt]{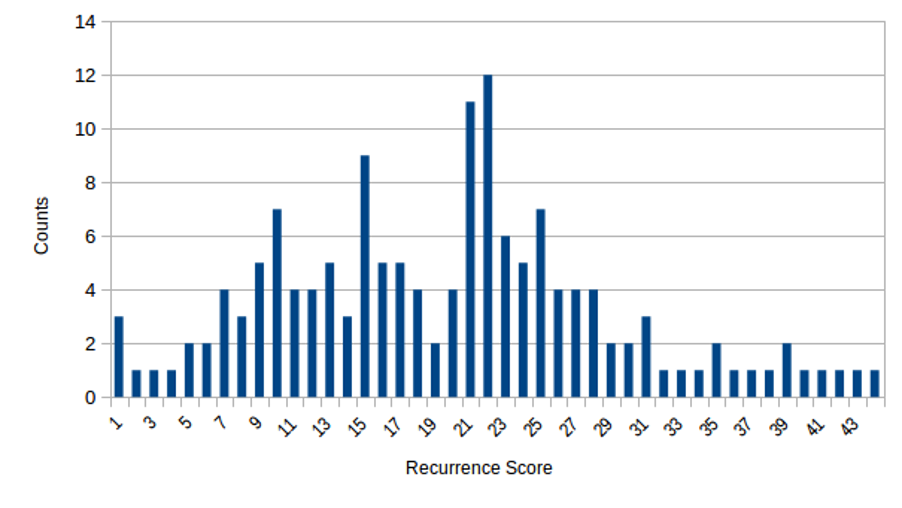}
    \caption{\textbf{Data histogram NCC and KUMC} \textit{It is a histogram of the RS score of data collected from two different sources. When RS is high, the histological grade is high, so most of them chose the system without the Oncotype DX test. Therefore, the number of data for the high score is small.}}
    \label{fig:figure1}
\end{figure} 

This study analyzed 125 hematoxylin and eosin (H\&E)-stained whole slide images (WSIs), each representing one of 125 unique patients diagnosed with early-stage breast cancer at two institutions: Korea University Medical Center Guro Hospital (KUMC) and the National Cancer Center (NCC). These WSIs were collected from surgeries performed between January 2014 and July 2019, excluding cases with prior chemotherapy or inadequate slide quality for histological analysis. Clinical information (e.g., age, ER/HER2 status, Ki67) was available for 47 patients who provided consent, as summarized in Table~\ref{tab:clinicopathologic}. All WSIs were scanned using an Aperio AT-2 scanner at ×20 magnification and divided into 512×512 patches. Otsu thresholding \cite{liao2001fast} was applied to remove background regions. Of 150 initial slides considered, 125 had corresponding Recurrence Score (RS) values from Oncotype DX and were included in the analysis, with their RS distribution illustrated in Figure~\ref{fig:figure1}. For patch-level analysis, images were labeled based on the following criteria: (1) patches with no cancer area were classified as benign; (2) patches with less than 25\% cancer area were also labeled benign; (3) remaining cancer patches were grouped according to the WSI’s RS value. This process yielded 245,544 benign patches and 103,031 cancer patches for examination at ×20 magnification.

\subsection{Model workflow}

\begin{figure}[hpt]
    \centering
    \includegraphics[width=360pt]{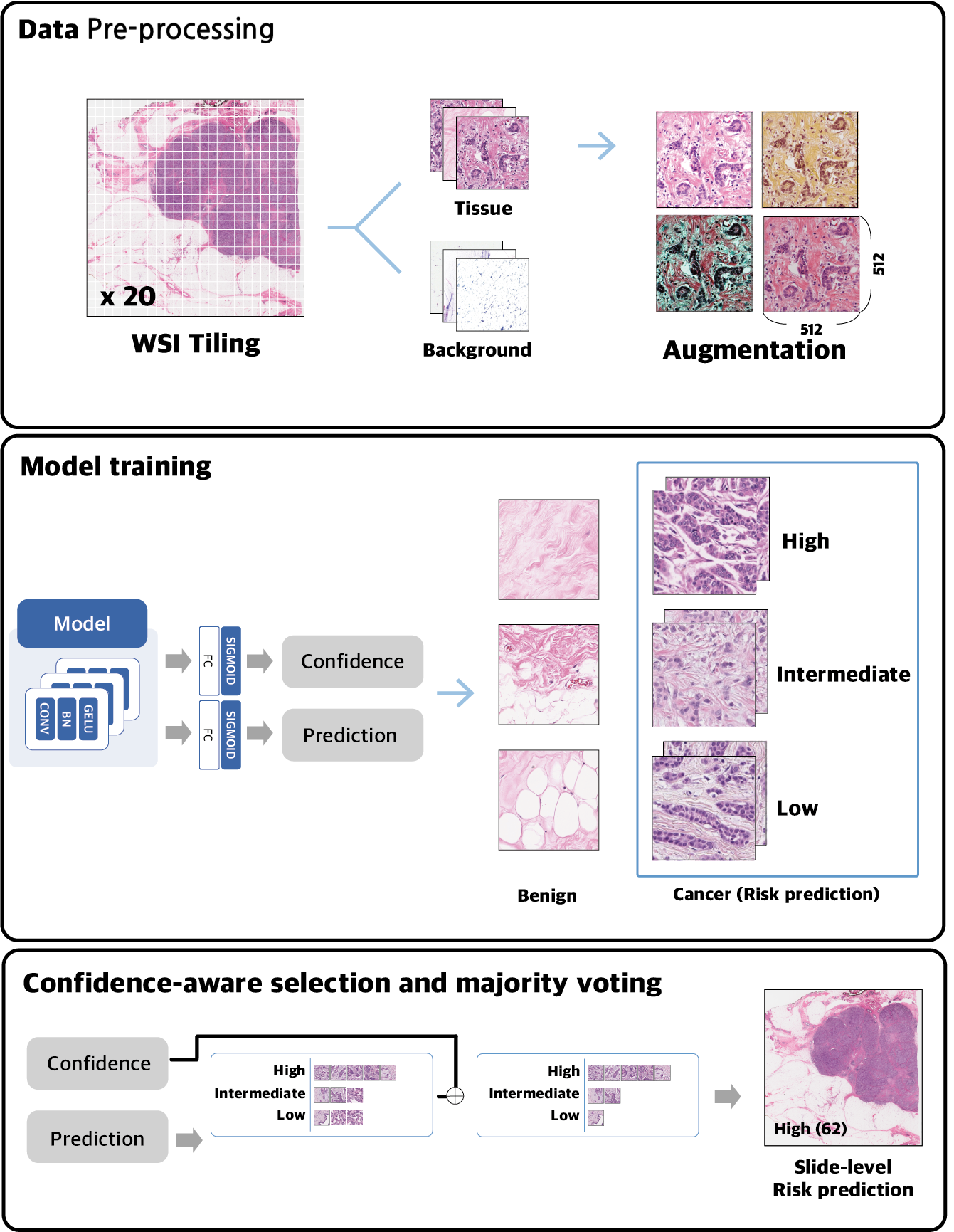}
    \caption{\textbf{Model workflow.} \textit{Whole Slide Images scanned at 200 magnification are tiled and classified into tissue and background. The image determined to be tissue is normalized through various methods of color augmentation and then entered as input into a model that checks for cancer. Individual cancer patches then predict each risk group and then vote on the risk group of WSI by referring to the confidence score.}}
    \label{fig:figure2}
\end{figure} 

Figure~\ref{fig:figure2} briefly shows how the model used in this study is learned and inference. First, when a hematoxylin and eosin stained whole slide image is input, image tiling is performed from the (0,0) point to the last point. For each tile (hereinafter referred to as a patch), a basic computer vision algorithm is used to first classify whether the patch is a background image or an image with tissue. In this study, Otsu thresholding \cite{4310076} was used for this purpose. Afterwards, normalization of the images is performed. This was done to stably learn and inference the model even when images of various dyeing environments came from various scanners. The distribution of images was increased using various augmentations such as gray scale, solarization, hue, and saturation. 
%\linenumbers

Afterwards, individual patches are entered into a patch-wise cancer classification model. Afterwards, the patches learn and classify whether cancer exists or not, and then enter the risk group classification model to predict high, intermediate, and low risk, respectively. After learning is completed in this way, the model selects patches to be used for WSI prediction based on confidence. After selection is completed, the final label of the WSI unit is predicted through majority voting of the selected patches.

\subsection{Model design and training scheme}
To conduct this study, eight NVIDIA RTX A5000 graphics cards were used. After tiling the images in patch units, the image normalization process was performed so that they could be generalized for various scanning and staining environments. To this end, various data augmentation methods were conducted to reduce color bias. For example, solarization, posterization, hue and saturation, and gray scale were applied. The generated patch is first entered into the model to determine whether it is cancerous. Patches identified as cancer predict recurrence risk groups according to RS values. For this learning, the ConvNext \cite{liu2022convnet} model pre-trained through ImageNet \cite{deng2009imagenet} was modified and used. Confidence-aware learning and patch-wise contrastive loss techniques were additionally applied to predict recurrence risk groups. We conducted a strict 5-fold cross-validation at the patient level. The dataset was partitioned into training, validation, and test sets, ensuring all patches derived from a single patient’s WSIs were exclusively assigned to one fold, thereby strictly avoiding data leakage. Additionally, each fold was stratified to maintain a consistent class distribution, approximating the ratio of low, intermediate, and high-risk cases observed in the original dataset (approximately 5:3:2)

\subsection{Confidence-aware learning}
In the case of deep learning, there is a  problem of over confidence \cite{moon2020confidence}. Over confidence refers to a phenomenon in which a model makes predictions with high probability even though the domain is a different image or an incorrect prediction. This causes the model's reliability to deteriorate. In particular, in this study, when predicting labels in units of WSI, majority voting, which makes predictions based on the most class among individual patches, is selected, so there is an effect caused by an unpredictable erroneous image. To overcome this, a measure called confidence was created, and the confidence of the actual prediction result of the model was additionally calculated.

\subsection{Patch-wise contrastive loss}
Chen et al., (2020)\cite{chen2020simple} proposed the contrastive loss function, which aims to map images that start from the same image in a vector space to a nearby space, even if the augmentation is different. Conversely, images that start from different images should be mapped to a far space. This function can be applied not only to the same image but also to images magnified in class units. Additionally, two different images with the same label should be farther away in vector space than images with different labels. \cite{lee2021supervised} observed that performance improved when this property was included in the model as a loss function. In this study, a contrastive loss was used between images with the same label in patch units to enforce this property.

\subsection{Loss function}
In this study, the cross entropy loss was basically used when predicting cancer status and prognosis. However, confidence-aware learning was performed to calculate how uncertain the patch was and to reflect it in the learning process of the model. If the actual model can extract a good visual representation vector that is helpful in predicting cancer or prognosis, the confidence level of the model will increase in proportion to the prediction result. If these properties are not trained correctly, the model will become overconfident in false results. In order to mitigate this overconfidence phenomenon to some extent, a method of continuous management of the confidence score derived by model training was adopted.

\begin{equation}
\label{eq1}
    l_{\text{ce}} = -\frac{1}{N} \sum_{i=1}^{N} \left( y_i \cdot \log(\hat{y}_i) + (1-y_i) \cdot \log(1-\hat{y}_i) \right)
\end{equation}

\begin{equation}
\label{eq2}
    % l_{\text{cf}} = \frac{1}{N} \sum_{i=1}^{N} (l_{\text{ce}_i} \cdot \text{CF}_i)
    l_{\text{cf}}(y_i, \hat{y}_i) = l_{\text{ce}}(y_i, \hat{y}_i) \cdot CF_i
\end{equation}

Equation \eqref{eq1} shows the cross entropy formula. In the case of $y$, it means the actual label, and in the case of $\hat{y}$, it means the predicted result of the model. Basically, if the predicted result of the model differs from the actual label, a loss occurs. At this time, individual cross entropy loss is multiplied by confidence scores according to Equation \eqref{eq2}.

\begin{equation}
\label{eq3}
\adjustbox{scale=0.9,center}{
$L_{\text{Reject Loss}}(i) =
\begin{cases}
\alpha \cdot l_{\text{cf}}(y_i, \hat{y}_i) + (1 - \alpha) \cdot l_{\text{ce}}(y_i, \hat{y}_i), & \text{if } CF_i \leq \text{threshold} \\
0, & \text{otherwise}
\end{cases}$
}
\end{equation}

Equation \eqref{eq3} is the reject loss created using Equations \eqref{eq1} and \eqref{eq2}. \cite{geifman2017selective} Basically, by adjusting the $\lambda$ term is hyper parameter, the number of rejections is determined according to the confidence score. In the case of $\alpha$, it is a term for which of the loss according to Equation \eqref{eq1} and the Equation \eqref{eq2} according to confidence should be given more weight. This confidence score is also used in the inference step, and is mainly used when determining the class of WSI. While predicting the probability of individual patches, if the confidence did not exceed $\lambda$, the threshold set during learning, the patch was defined as a patch that was not helpful when predicting WSIs.

\begin{align}
\label{eq4}
L_{\text{Contrastive}} = & -\log\left(\frac{\exp(z_{i} \cdot z_{j} / \tau)}{\exp(z_{i} \cdot z_{j} / \tau)}\right) \nonumber \\
&+ \alpha_{\text{pos}} \cdot \sum_{k=1}^{2N} \mathbbm{1}[k \neq i] \mathbbm{1}[y_{k} = y_{i}] \wedge P_{\mathrm{wsi}} \exp(z_{i} \cdot z_{k} / \tau) \nonumber \\
&+ \alpha_{\text{neg}} \cdot \sum_{k=1}^{2N} \mathbbm{1}[k \neq i] \mathbbm{1}[y_{k} \neq y_{i}] \wedge N_{\mathrm{wsi}} \exp(z_{i} \cdot z_{k} / \tau)
\end{align}

Equation \eqref{eq4} is based on contrastive loss \cite{chen2020simple}. Contrastive loss is a method often used in self-supervised learning, and it is a theory that images from oneself should be located at a short distance in the vector space, and images from other images should be located at a long distance in the vector space. In the formula, $z$ means an individual patch, and $y$ means the prognosis wise label of the patches. In the case of $\alpha$, it is a hyper-parameter that means the weight for each positive and negative. According to the above formula, all patches must be close to each other in the vector space with the same WSI, and at the same time, the positions of patches with the same class label must be close.

\begin{equation}
\label{eq5}
L_{\text{total}} = L_{\text{rejectloss}} + \psi_{\text{contrastive}} \cdot L_{\text{contrastive}}
\end{equation}

Equation \eqref{eq5} is the final loss function. $\psi$ is a weight for contrastive loss, and it is designed in a form that is set to 0 at the beginning of learning for reject loss and gradually increases to 1.0 as the epoch progresses.

\section{Results}
The clinicopathologic characteristics of the study cohort are summarized in Table~\ref{tab:clinicopathologic}.
Due to patient privacy constraints and institutional regulations, only limited anonymized clinical data were accessible.
A total of 125 patients were included in this study, along with 125 whole-slide images (WSIs) deemed significant by pathologists. Among these, clinical information was available for 47 patients who consented to data sharing. The mean patient age was 56 years (range, 37–79 years), and Ki67 data were available for 46 out of the 47 cases.
We conduct both quantitative and qualitative analyses of the model's outcomes. Based on the widely recognized Genomic Health Inc. Recurrence Score (GHI-RS) criteria for risk categorization, we designated the risk levels as low (scores below 18), intermediate (scores between 18 and 31), and high (scores above 31).
The key findings of this investigation are outlined below:
The model demonstrated a 91.2\% accuracy rate in identifying high-risk cases from WSIs. The composite accuracy for categorizing cases into high, intermediate, and low risk was approximately 87.33\%.
When analyzing individual image patches, the overall accuracy reached 87.75\%. Notably, the model did not misclassify any high-risk images as low risk.

\begin{table}[htbp]
\centering
\renewcommand{\arraystretch}{1.2}

\label{tab:clinicopathologic}
\begin{tabular}{l c}
\hline
Characteristic & Value \\
\hline
Total patients, n                      & 47 \\
Age, years, mean (range)               & 56 (37–79) \\
Hormone receptor status, n (\%)        & \\
\quad ER positive                      & 47 (100\%) \\
\quad HER2-negative or equivocal       & 47 (100\%) \\
Histologic grade, n (\%) & \\
\quad Grade 1                          & 32 (68.1\%) \\
\quad Grade 2                          & 15 (31.9\%) \\
Tumor size, cm, mean (range)           & 1.08 (0.6–2.5)\\
Ki67, mean (\%), (range)               & 11.4 (0.9–33.8) \\
\hline
\end{tabular}
\caption{Clinicopathologic characteristics of the study cohort (n=47). 
All patients were female, hormone receptor-positive (ER-positive), and HER2-negative or equivocal. Ki67 data were partially available (n=46).}
\end{table}

\subsection{Quantitative Analysis}
We quantitatively evaluate the model's performance. Due to the large size of Whole Slide Images (WSIs), directly learning from them poses challenges, particularly in accurately learning morphological features. To address this, patches were extracted from WSIs using a tiling process for input into the model. For the purpose of training the classification model, labels corresponding to the WSI level were applied to each patch. This approach considered the source WSI of each patch without taking into account specific prognostic criteria for breast cancer, such as tumor tubule formation, nuclear pleomorphism, and mitotic figure count. The creation of approximately 170,000 patches following this method yielded results consistent with those presented in Table~\ref{tab:patch_level_table_1}. The overall accuracy for the patch-based predictions was 87.75\%, with the final WSI outcomes derived from these patch predictions detailed in Table~\ref{tab:patch_level_table_1}.

\begin{table}[]
\centering
\renewcommand{\arraystretch}{1}
\begin{tabular}{l|c|c|c|c}
\hline
\multicolumn{1}{c|}{} & \multicolumn{4}{c}{Prediction} \\

\multirow{-2}{*}{Model} & Benign & Low & Intermediate & High \\
\hline
Benign          & 30090 & 227   & 1033  & 180 \\
Low             & 137   & 725   & 259   & 0   \\
Intermediate    & 1585  & 1583  & 15268 & 1853 \\
High            & 9     & 0     & 192   & 4474 \\
\hline
\end{tabular}
\caption{Direct patch level classification model: The diagonal values are number of correct patches in deep learning model’s with total accuracy of 87.75\%.}
\label{tab:patch_level_table_1}
\end{table}

\begin{table}[]
\centering
\renewcommand{\arraystretch}{1.2}
\begin{tabular}{c|c|c|c|c}
\hline
\multirow{2}{*}{True Label} & \multicolumn{4}{c}{Model Prediction} \\

 & Low & Intermediate & High & Total \\
\hline
Low          & 42 & 5  & 2  & 49 \\
Intermediate & 14 & 44 & 1  & 59 \\
High         & 0  & 8  & 9  & 17 \\
\hline
Total        & 56 & 57 & 12 & 125 \\
\hline
\end{tabular}
\caption{Confusion Matrix of Breast cancer Recurrence score.}
\label{tab:Recurrence score confusion}
\end{table}

The process for predicting outcomes utilizing the model is explained as follows: The model first analyzes tiled patches derived from Whole Slide Images (WSIs). Secondly, it determines if the input patch shows cancer tissue. Thirdly, it computes the likelihood of breast cancer recurrence for patches identified as cancer. Table~\ref{tab:Recurrence score confusion} displays real risk categories based on the 21-gene test results in the rows and the model's predictions in the columns. The analysis covered 125 images, and notably, there were no instances where patients at high risk were incorrectly classified as low risk. Table~\ref{tab:my-table3}provides a breakdown of the model's performance using a confusion matrix. The sensitivity scores were not ideal, but the model showed high specificity.

\begin{figure}
    \centering
    \includegraphics[width=360pt]{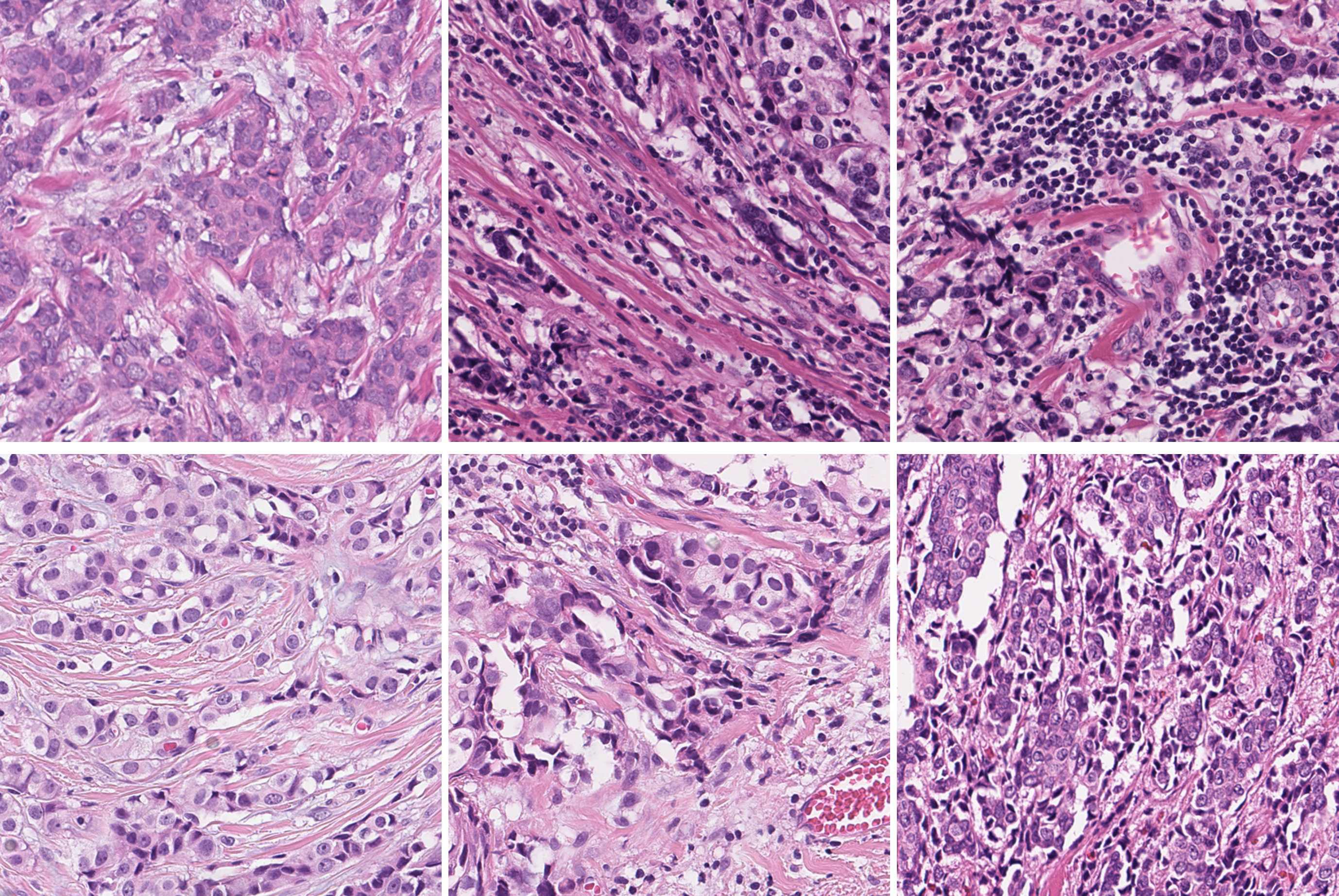}
    \caption{Patch-level examples shows misclassification cases resulting from naive majority voting. 
    \textbf{Top row:} Example patches from a high-risk case misclassified as intermediate-risk at the WSI-level, although most patches were individually predicted correctly as high-risk. 
    \textbf{Bottom row:} Example patches from a low-risk case misclassified as intermediate-risk at the WSI-level, despite accurate low-risk predictions at the patch-level. 
    These results highlight the limitations of using naive majority voting for WSI-level predictions, suggesting potential improvements through more sophisticated patch aggregation methods.
    }
    \label{fig:figure_add}
\end{figure}

We further investigated cases of misclassification and identified an important limitation of our current aggregation method. Several misclassified WSIs contained patches that were accurately classified when evaluated individually. Specifically, multiple patches within intermediate-risk WSIs incorrectly classified by majority voting were individually and accurately classified as high-risk at the patch level, demonstrating that the model successfully captured the relevant histological patterns indicative of high recurrence risk at the local tissue level (Figure~\ref{fig:figure_add}). These inconsistencies predominantly arose due to the simplistic majority voting mechanism applied at the WSI level, which failed to incorporate patch-level prediction confidence or clinical significance. Future investigations could explore the adoption of more advanced patch aggregation methods, such as weighted voting or attention-based strategies, to better leverage the model’s accurate predictions at the patch level.

\begin{table}[]
\centering
\renewcommand{\arraystretch}{1.2}
\begin{tabular}{l|c|c|c}
\hline
\multicolumn{1}{c|}{} & \multicolumn{3}{c}{WSI level Performance} \\

\multirow{-2}{*}{} & Low RS & Intermediate RS & High RS \\
\hline
Accuracy    & 0.932 & 0.776 & 0.912 \\
Sensitivity & 0.857 & 0.746 & 0.529 \\
Specificity & 0.816 & 0.803 & 0.972 \\
PPV         & 0.851 & 0.772 & 0.750 \\
\hline
\end{tabular}
\caption{WSI level Performance metrics.}
\label{tab:my-table3}
\end{table}

\subsection{Qualitative Analysis}

We conduct a qualitative assessment of the model's performance through the following steps: Firstly, we ensure that the model's performance is not biased by the number of patches viewed, preventing any class dominance within the Whole Slide Images (WSIs). Secondly, we examine the distribution of the Recurrence Score (RS) values predicted by the model to ascertain their meaningfulness.

\begin{figure}[htp]
    \centering
    \includegraphics[width=240pt]{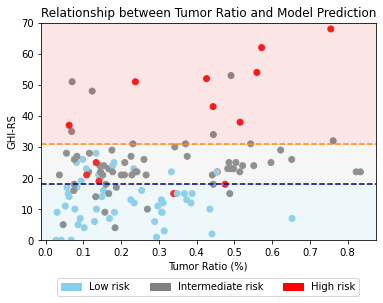}
    \caption{\textbf{Relationship between cancer area ratio and model prediction.} 
 Except for one exception, in most cases, it was confirmed that when the cancer area was large (accounting for more than 50\% of the total area), the GHI-RS was generally high. Additionally, even in the two cases where the model misclassified (the two red points below the navy-colored dashed line), it can be seen that the cancer area takes up more than 30\% of the entire image.
}
    \label{fig:figure3}
\end{figure} 

Figure~\ref{fig:figure3} presents a graph showing the relationship between the proportion of cancerous area within the total image and the GHI-RS values. The x-axis indicates the extent of cancer presence relative to the entire image, while the y-axis displays the GHI-RS values. In this illustration, a navy dotted line marks the low-risk threshold as defined in this study, whereas a dark orange dotted line denotes the high-risk threshold. Each point on the graph represents a model prediction. Contrary to what might be expected.

\begin{figure}[htp]
    \centering
    \includegraphics[width=380pt]{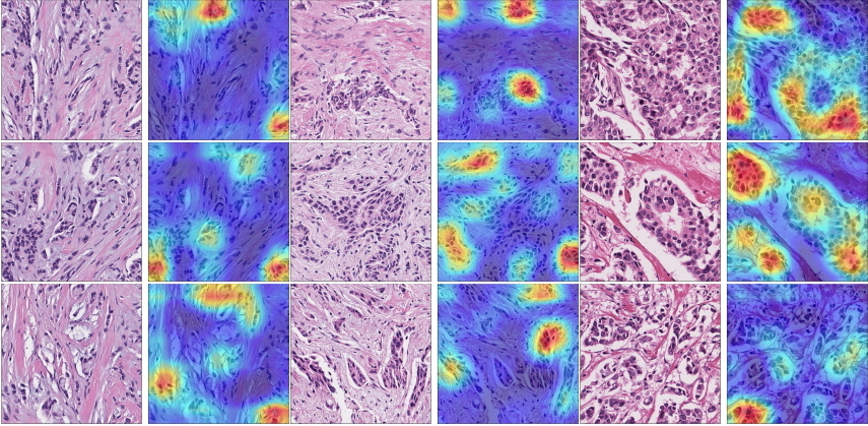}
    \caption{\textbf{Corrected Grad-CAM Images.} \textit{In order from left to right, it shows low, intermediate, and high risk images and Grad-CAM for the corresponding images, respectively. Individual Grad-CAM images are drawn in the form of a heat map, so that if the actual predicted value is high, the color is red, and if it is low or absent, the color is close to blue.}}
    \label{fig:figure4}
\end{figure} 

\begin{table}[]
\centering
\begin{tabular}{cc|ccc}
\hline
\multirow{3}{*}{\begin{tabular}[c]{@{}c@{}}Nottingham\\ Histology\\ Grade\end{tabular}} & Grade 1 & 17077 & 3909 & 733 \\
 & Grade 2 & 4802    & 34114           & 10779    \\
 & Grade 3 & 27      & 4500            & 10621    \\ \hline
 &         & Low     & Intermediate    & High     \\
 &         & \multicolumn{3}{c}{Model Prediction} \\ \hline
\end{tabular}
\caption{Confusion matrix for nottingham grade and model prediction results. For low risk (Grade 1), sensitivity 0.78 and specificity 0.92 are obtained. For intermediate risk (Grade 2), sensitivity 0.68, specificity 0.77. For high risk (Grade 3), sensitivity is 0.70 and specificity is 0.83.}
\label{tab:table4}
\end{table}

\begin{figure}[htp]
    \centering
    \includegraphics[width=360pt]{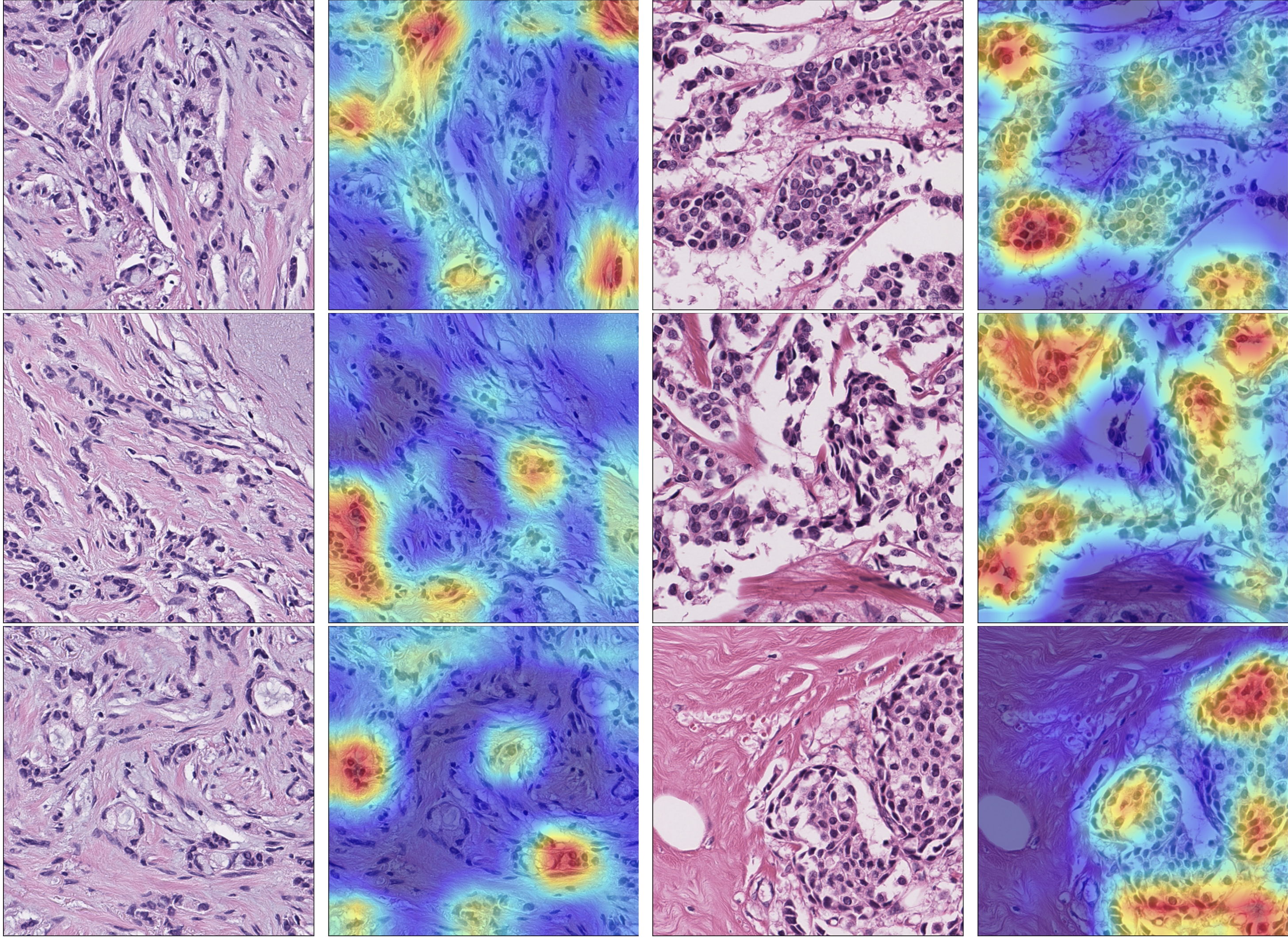}
    \caption{This visualization shows Grad-CAM results comparing Grade 1 (low risk) on the left and Grade 3 (high risk) on the right. The left column (low risk) indicates that the model primarily based its prediction on areas where tubular structures were preserved, regions with uniform nuclear morphology, and connective tissue areas. The model's response in these regions is weak, as indicated by the predominance of blue in the Grad-CAM heatmap, suggesting that the model did not consider these areas significant. On the other hand, in the right column (high risk), the model strongly identified features such as the disruption or loss of tubular structures, irregular nuclear dysmorphism, high cell density, and increased mitotic activity. As a result, the Grad-CAM heatmap highlights these regions in red and yellow, indicating that the model considered these features crucial for high-risk classification. These findings align with pathological assessments, suggesting that the model effectively differentiates between low- and high-risk cases by utilizing tissue structure preservation, nuclear morphology changes, and cell density as key pathological features.}
    \label{fig:figure5}
\end{figure}

\begin{figure}[htp]
    \centering
    \includegraphics[width=240pt, height=240pt]{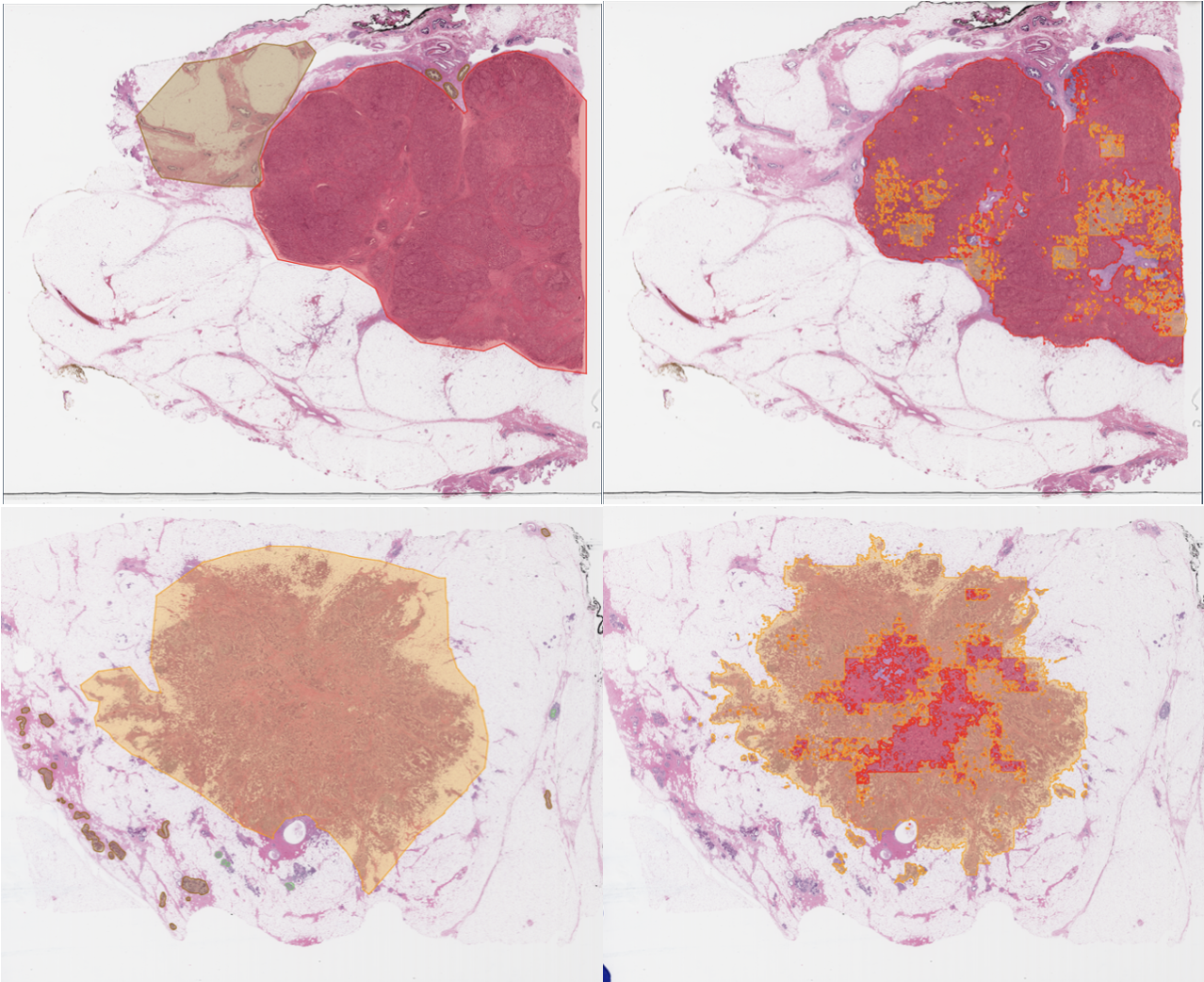}
    \caption{Comparison of model prediction results and pathologist opinions. \textit{Comparison of model prediction results with WSI unit results of pathologists. The results of grade 1 (low risk), grade 2 (intermediate risk), and grade 3 (high risk) are shown in the order of yellow, orange, and red, respectively. Since the prediction of the model is majority voting, visualization was performed for a case where the correct answer for each side risk was correct.}}
    \label{fig:figure6}
\end{figure}

Figure~\ref{fig:figure4} presents a class activation map to illustrate how the model identifies morphological features. Each image within the figure utilizes a heatmap to indicate the areas the model has identified as belonging to a specific class. Areas highlighted in red signify sections where the model found strong evidence for its classification. The figure pairs each original image with its corresponding class activation map: the first column displays images classified as low risk, the second column includes those deemed intermediate risk, and the third column contains images classified as high risk.
Table~\ref{table4} shows the confusion matrix for these domain-level annotations. Sensitivity 0.78 and specificity 0.92 came out for low risk (Grade 1). It was confirmed that 0.68 and 0.77 for Intermediate risk (grade 2) and 0.70 and 0.83 for High risk (Grade 3). (Correlation 0.61 for region of interesting, correlation excluding Intermediate (Grade2) was 0.79)
Figure~\ref{fig:figure5} and Figure~\ref{fig:figure6} are shows. The results of the model and the pathologist's generally agree. As a result of visualizing these images through CAM, it was confirmed that highlight was actually applied to the cancer area. In the case of Figure~\ref{fig:figure6}, the results of the model and the opinions of the pathologist were compared in WSI units. For individual cases, the GHI-RS score is 62 points (high risk) for the image above and 22 points (intermediate risk) for the image below. 

\section{Discussion}

In the current investigation, we conduct a comparative analysis between deep learning algorithms and the established multigene assay, Oncotype DX, for the purpose of profiling breast cancer risk. The study reveals an 87.75\% concordance rate in stratifying patients into low, intermediate, and high-risk categories between these two distinct methodologies. This substantial concordance aligns with previous findings on the consistency across various multigene assays, underscoring the methodological robustness of both approaches. Specifically, within the context of the OPTIMA Prelim trial, the agreement between risk classifications by Prosigna and Oncotype DX Recurrence Score (RS) (Exact Sciences, Madison, USA) was observed at 81.0\% for the low and high-risk groups, and 77.5\% for a combined assessment of low/intermediate and high-risk groups. Such variance is to be expected due to the inherent differences in the underlying models, gene sets, and clinical variables utilized by each assay. Therefore, a concordance of 87.75\% is considered to be a very high level.

Deep learning models have a problem of overconfident. \cite{mehrtash2020confidence} However, this excessive confidence reduces the reliability of the model in the area of medical image analysis. In this study, we tried to secure the reliability of the model while alleviating the problem of overconfident of this model to some extent.
Many previous studies have not paid attention to this overconfident problem even when predicting recurrence risk (RS). The study of Kim et al., (2021)\cite{kim2021deep} was conducted by inputting information such as age, DCIS size, size of lymph node metastasis, size of tumor, Ki-67 status, and histological grade, so that the RS group was divided into low and high groups. Although this was predicted with relatively high accuracy, there were some regrets in specifically confirming which features acted importantly. In the case of the study of Baltres et al., (2020)\cite{baltres2020prediction}, GHI-RS was predicted using information such as age and tumor size histological type. Compared to other studies, it was successfully performed to analyze the correlation between each data, but the performance of the individual risk group was not significantly good compared to other studies.
Similarly, in the study of Williams et al., (2022)\cite{williams2022use}, the work of predicting the RS group was performed by using information such as tumor size, histology information, tumor grade, and lymphovascular invasion. This could show a compliant performance of 97\% specificity, but likewise, the correlation between features was not considered at all. The study of Pawloski et al., (2022)\cite{pawloski2022supervised} also compared correlations such as ER, PR, and HER2, but did not analyze why such insights could be obtained.
The purpose of this study was to prove that there is a histology grade relationship with gene test results, alleviating the problem of overconfidence in deep learning models to some extent. First, the reject loss allowed us to remove patches that were difficult to predict or images that were very different from patches of the same class. Second, experiments through comparison with domain-level annotations by experienced pathologists were an important clue to confirming the reliability of the model. This experiment was conducted based on the following assumptions: a) The 21-gene test is a method to check the expression level of 21 genes, including molecular subtype status such as ER and HER2. b) Molecular subtype status affects the growth and proliferation of cancer. c) The histological grade of an experienced pathologist is determined based on the growth of cancer. d) Therefore, if the model learned morphological characteristics well from the data, it would show similar results to the histological grade of experienced pathologists. To confirm this, we received territorial unit annotations for all 125 images through experienced specialists. As detailed as possible, grade was made including tubular formation, nuclear pleomorphism, and mitotic count in units of areas, and we compared it with the results of the model. Based on this, this study confirmed that the genetic test results can be approximated using WSI stained with H\&E in the limited situation of the deep learning model. We also confirmed to what extent the criteria used by pathologists to determine tissue ratings by region annotation and the patterns learned by the deep learning model match. 
\begin{figure}[htp]
    \centering
    \includegraphics[width=240pt]{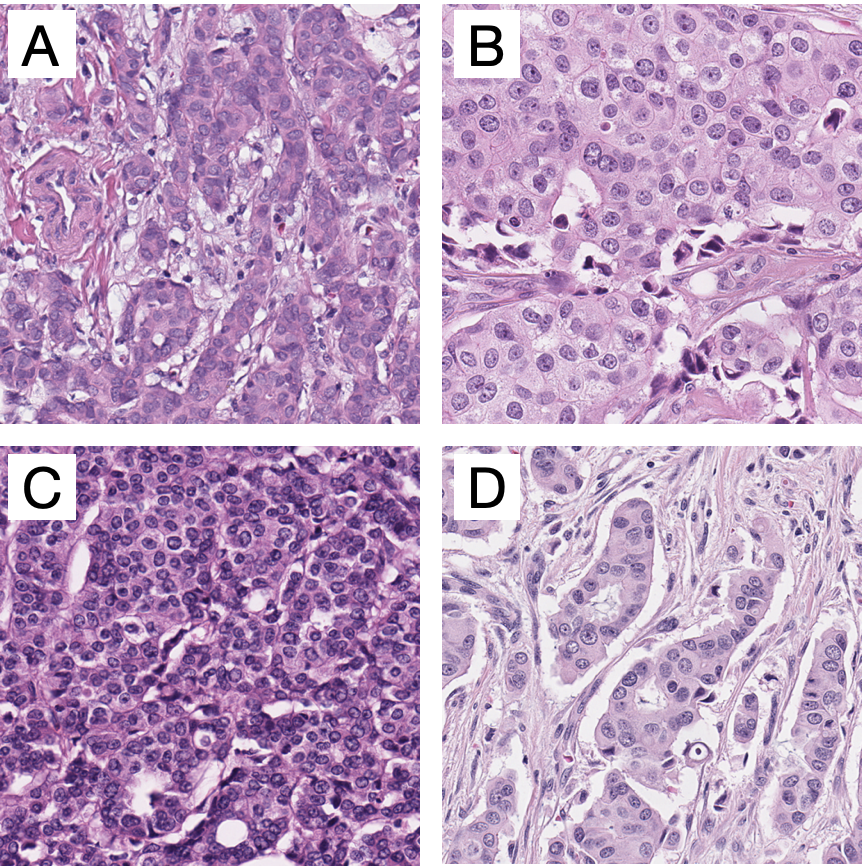}
    \caption{When the pathologist's opinion does not match the RS group, the model prediction adjusts the rs group. \textit{(A) and (C) are images corresponding to the low risk group, respectively, and (B) and (C) are images corresponding to the high risk group. The model precisely matched the group for the corresponding images, but the pathologist's opinion produced the opposite result.}}
    \label{fig:figure7}
\end{figure} 
Our results confirm that, under the limited circumstances of deep learning models, H\&E-stained WSI can be used to approximate the genetic test results. Furthermore, we confirm that the criteria used by pathologists for judging tissue ratings by region annotation and the patterns learned by deep learning models are consistent to some extent. As shown in Figure~\ref{fig:figure7}, when the pathologist's opinion does not match the RS group, the model's prediction adjusts the risk score group. This figure illustrates specific cases where the model correctly classified the images into low or high-risk groups, even when the pathologist's assessment was contradictory.
Unfortunately, there are several limitations. Typically, the model works well for low and high-risk data, but struggles with unusual data points like intermediate risk or those with a recurrence score of 0 or 44 and above, leading to bias or difficulty in assessment. Secondly, certain instances were entirely omitted during data collection, including those deemed pathologically meaningless, and data from other institutions were disregarded. Nevertheless, it was confirmed that the deep learning model can partially imitate histological properties from a limited amount of data. In the future, with enough data collected, it is believed that more research can advance this. Lastly, it is important to acknowledge that the observed correlation between our model’s predictions and expert histological grading (Pearson correlation coefficient: 0.61) is both expected and clinically meaningful. The Oncotype DX recurrence score itself is known to significantly correlate with traditional histological parameters such as tumor grade and proliferation markers \cite{paik2004multigene, sparano2018adjuvant}. Thus, the similarity in model predictions and pathologists’ assessments suggests that our approach effectively captures clinically relevant morphological features. However, further studies integrating additional biomarkers, such as Ki-67, and utilizing larger, more balanced datasets are needed to better clarify the incremental prognostic value offered by our deep learning-based method beyond traditional pathological assessment. Although promising, our study utilized a relatively small cohort of 125 patients from two institutions, with an inherent imbalance in recurrence risk categories, particularly a shortage of high-risk cases. Clinical information was limited to 47 consented cases, further constraining detailed clinicopathologic correlations. Therefore, further validation with larger, balanced datasets is essential to establish the clinical utility of this deep learning approach.

%This study used limited data, except for patients who received preoperative chemotherapy or H\&E slides that were pathologically unsuitable for clinical trials. The goal was to evaluate how well AI-based software responded to H\&E slide reading results and molecular genetic prognostic factors (Oncotypes Dx) results in breast cancer tissues. This experiment was based on limited data and annotations.

\section{Conclusion}

Through this study, we were able to show that even with only H\&E stained WSI, we can approximate the gene test results to some extent. In addition, it was confirmed that deep learning models can learn morphological characteristics such as mitotic figures through various visualization methods.
Deep learning models generally tended to follow pathologist's opinion, but they were not completely consistent.
Through the results of the above studies, it was found that histological grading could be used to approximate this to some extent, and if we expand it further, we expect to be able to effectively prove recurrence by using common patterns in breast cancer.

\section*{Declarations}

\subsection*{Ethical Approval and Consent to Participate }
This study was reviewed and approved by the Institutional Review Board (IRB) of Korea University Medical Center, Seoul, Korea (IRB No. 2020-GRO-146). This study was performed in accordance with the Declaration of Helsinki. Written informed consent for participation was not required for this study in accordance with the national legislation and the institutional requirements.

\subsection*{Consent for Publication}
Not applicable 

\subsection*{Conflicts of Interest}
Kim, S. is the CEO of Deep Bio Inc., Kwak, T. is the CTO, and Chang, H. is the Medical Officer. Lee, J. and Lee, G. are employees of Deep Bio Inc. Kwon, Y. and Kim, C. have no conflicts of interest to declare.

\subsection*{Availability of Supporting Data}
The data that support the findings of this study are available from Korea University Guro Hospital and National Cancer Center but restrictions apply to the availability of these data, which were used under license for the current study, and so are not publicly available. Data are however available from the authors upon reasonable request and with permission of Korea University Guro Hospital and National Cancer Center. Interested researchers should contact the corresponding author, Kim, C. at idea1@korea.ac.kr 

\subsection*{Funding}
This research received no external funding.

\subsection*{Authors’ Contributions}
G.L. conducted the overall experiments and wrote the manuscript. J.L., T.K., S.K., and H.C. contributed to manuscript revision and proofreading. Y.K., H.C., and C.K. provided expert pathological insights, contributed to data collection, labeling, and overall interpretation. S.K. assisted in setting up the experimental framework and necessary configurations. All authors reviewed and approved the final manuscript.

\subsection*{Acknowledgements}
We sincerely thank Deep Bio Inc., Korea University Guro Hospital, and the National Cancer Center (NCC) for their support in conducting this research.

\bibliography{sn-bibliography}% common bib file

\end{document}